\documentclass[pra,twocolumn,aps,showpacs,superscriptaddress]{revtex4-1}

\usepackage{amsmath}
\usepackage{amssymb}
\usepackage{graphicx}
\usepackage{subfigure}
\usepackage[usenames,dvipsnames]{color}
\usepackage{setspace}

\begin{document}

%%%%%%%%%%%%%%%%%%%%%%%%%%%%%%%%%%%%%%%%%%%%%%%%%%%%%%%%%%%%%%%%%%%%%%%%%%%%%%
%%%%                     Title and authors                                %%%%
%%%%%%%%%%%%%%%%%%%%%%%%%%%%%%%%%%%%%%%%%%%%%%%%%%%%%%%%%%%%%%%%%%%%%%%%%%%%%%

\title{Crow instability in unitary Fermi gas}

\author{S. Gautam} 
\affiliation{Indian Institute of Science,
             Bangalore - 560 012, India}

%%%%%%%%%%%%%%%%%%%%%%%%%%%%%%%%%%%%%%%%%%%%%%%%%%%%%%%%%%%%%%%%%%%%%%%%%%%%%%
%%%%%%%%%%                    Abstract                             %%%%%%%%%%%
%%%%%%%%%%%%%%%%%%%%%%%%%%%%%%%%%%%%%%%%%%%%%%%%%%%%%%%%%%%%%%%%%%%%%%%%%%%%%%

\date{\today}
\begin{abstract}
We investigate the initiation and subsequent evolution of Crow instability 
in an inhomogeneous unitary Fermi gas using zero-temperature Galilei-invariant
non-linear Schr\"odinger equation. Considering a cigar-shaped unitary Fermi
gas, we generate the vortex-antivortex pair either by phase-imprinting or
by moving a Gaussian obstacle potential. We observe that the Crow instability 
in a unitary Fermi gas leads to the decay of the vortex-antivortex pair into multiple 
vortex rings and ultimately into sound waves.   
\end{abstract}


\maketitle

%%%%%%%%%%%%%%%%%%%%%%%%%%%%%%%%%%%%%%%%%%%%%%%%%%%%%%%%%%%%%%%%%%%%%%%%%%%%%%
%%%%%%%%%%%%%                 Introduction                         %%%%%%%%%%%
%%%%%%%%%%%%%%%%%%%%%%%%%%%%%%%%%%%%%%%%%%%%%%%%%%%%%%%%%%%%%%%%%%%%%%%%%%%%%%

\section{Introduction}
\label{I}
A vortex-antivortex pair in a three dimensional inviscid fluid is susceptible to 
the long wavelength
cooperative instability known as Crow instability (CI) \cite{crow.8.2172}.
In an undisturbed vortex-antivortex pair, each vortex feels the velocity field
produced by the other. As a result of it, the pair moves in the direction
perpendicular to the plane containing the pair. Now, some initial disturbance 
in the fluid can lead to the formation of the sinusoidal distortions 
in the vortex shape, which are symmetric with respect to the plane separating the
two vortices. In this case, the total velocity field experienced by the
vortex (antivortex) has three contributions: (a) velocity field induced by 
the vortex (antivortex) on itself due to its curvature, (b) the straining experienced 
by the vortex (antivortex) when displaced from its equilibrium position in 
the velocity field induced by the undisturbed antivortex (vortex), and (c) the 
velocity field induced on the vortex (antivortex) by the sinusoidally distorted
antivortex (vortex). The strain rate is directly proportional to circulation of
the vortex and inversely proportional to square of the distance between the vortices.
The first and the sum total of the last two contributions 
are also termed as the velocity field produced due to the self and mutual-induction
respectively. In the absence of the second vortex and hence mutual-induction, 
the sinusoidal displacement disturbance travels along the vortex line
and rotates about its unperturbed position, thus distorting the vortex line
into a helical shape. These  helical waves, which are neutrally stable, propagating 
along the length of the vortex line are also known as Kelvin waves \cite{Kelvin}. In the presence of 
the second vortex, the self induced rotation can be decreased or increased due 
to the azimuthal velocity component of strain field. When the two balance each 
other, the unbalanced radial component of the strain field leads to the exponential 
growth of the displacement perturbation; this is known as CI. As the amplitude of 
these sinusoidal oscillations reaches a critical value, the vortex and antivortex 
reconnect leading to the formation of a chain of vortex rings. It was shown by Crow
that a vortex line in the presence of only strain field will be unconditionally unstable,
whereas it will be neutrally stable of only self induction were present \cite{crow.8.2172}. 
The interplay between these two opposing effects leads to the CI.
The CI is also 
responsible for the decay of wing-tip generated vortex-antivortex pair in the wake of
the aircraft \cite{crow.8.2172}. The CI in this case can be triggered by the
turbulence in surrounding air or local variation in the air temperature or
density. It was shown by Kuznetsov et al. \cite{PhysRevE.51.4479}, that  
only the long-wavelength symmetric 
modes are unstable due to CI. The long-wavelength antisymmetric modes, on the other
hand, are stable \cite{PhysRevE.51.4479}. The typical wavelength of the unstable modes is 
much larger than the size of the vortex core \cite{crow.8.2172}.

The CI can occur not only in classical fluids but also in 
superfluids. The characteristic feature of the superfluids is the quantized 
circulation in them \cite{onsager.6.279,feynman1955}. Hence, the realization 
of quantized vortices in Bose-Einstein condensates (BECs) 
\cite{PhysRevLett.83.2498,PhysRevLett.84.806,Science.292.476} and 
two-component Fermi superfluids \cite{Nature.332.1288} has established the 
superfluid nature of these systems.

In case of homogeneous BECs, the CI has been 
investigated theoretically in Refs.~\cite{PhysRevE.51.4479,JPhysA.34.10057}.
In this context, Kuznetsov et al., \cite{PhysRevE.51.4479} demonstrated the 
long wavelength instability of acoustic solitons and vortex pairs in the 
frame work of three dimensional Schr\"odinger equation. It was noted in this 
study that in case of solitons, the instability corresponds to snake 
instability \cite{kadomtsev.192.753}, while in case of vortex-antivortex pairs
it takes the form of CI. Recently, CI in cigar shaped BECs was theoretically 
studied in Ref.~\cite{PhysRevA.84.021603}. On the experimental front, the 
snake instability of the solitons leading to the formation of vortex rings has
been observed \cite{PhysRevLett.86.2926,Science.293.663}, whereas the CI is 
still eluding the experimental observation. The CI is inhibited in oblate 
BECs. This is due to the suppression of bending and reconnections of 
vortices in oblate BECs \cite{PhysRevLett.104.160401}. Also, the generation of 
Kelvin waves \cite{PhysRevLett.90.100403, PhysRevLett.101.020402}, which aids 
the development of the CI, is suppressed in oblate condensates 
\cite{PhysRevA.84.023637}. 

 In case of two-component Fermi gas, the attractive interaction between the
two component spin states is a prerequisite for superfluidity. Depending upon 
the value of $s$-wave scattering length and Fermi wave vector $k_{\rm F}$, 
these superfluids can be weakly interacting $k_{\rm F}|a|\ll1$ or strongly 
interacting $k_{\rm F}r_{\rm e}\ll1\lesssim k_{\rm F}|a|$, where $r_{\rm e}$
is the effective range of interatomic interaction potential. The two limiting cases, 
$k_{\rm F}|a|\rightarrow 0$ 
($a\rightarrow-0$) and $k_{\rm F}|a|\rightarrow\infty$ 
($a\rightarrow -\infty$) are known as Bardeen-Cooper-Schrieffer (BCS) and 
unitary limits respectively. The inverse of Fermi wave vector $k_{\rm F}$
is of the order of average interparticle separation. Thus, the effective
range is much less than the interparticle separation which in turn is much less 
than $s$-wave scattering length in the unitary limit. It implies that the unitary Fermi gas 
(UFG) is simultaneously dilute ($r_{\rm e}\ll k_{\rm F}^{-1}$) and strongly 
interacting ($k_{\rm F}^{-1}\ll|a|$).
In the unitary Fermi gas (UFG), the 
thermodynamic quantities become the universal functions of Fermi energy 
$E_F$ and the ratio of temperature to Fermi temperature $T/T_{F}$
\cite{RevModPhys.80.1215}. The generation, real time evolution, and 
interaction among the quantized vortices has been studied in UFG using the 
time-dependent superfluid local density approximation at zero-temperature 
\cite{bulgac2011real}. The method involves the computationally intensive task 
of solving time dependent coupled Bogoliubov-de Gennes equations 
self-consistently. An alternative method based on the single orbital time 
dependent density functional theory has been developed in 
Refs. \cite{PhysLettA.327.397,PhysRevA.70.033612,PhysRevA.71.033625,
PhysRevA.73.065601,salasnich2007mean,EuroPhysLett.79.50003,
PhysRevA.76.043626,PhysRevA.77.043609,PhysRevA.78.053626,PhysRevA.78.043616,
PhysRevA.77.045602,PhysRevA.79.023611,JPhysB.42.215306,NewJPhys.11.023011,
LaserPhys.4.642} to study the superfluid Fermi gases. Employing this approach,
the UFG is well described by a non-linear Schr\"odinger equation (NLSE) at 
$T=0$K, which leads to same superfluid density as the original many body 
system. The NLSE is appropriate to study UFG, provided the characteristic 
wavelength of the phenomenon under study is larger then the coherence length 
\cite{LaserPhys.4.642}. The NLSE has been used to study various phenomena like
collective modes \cite{PhysLettA.327.397, PhysRevA.70.033612, 
PhysRevA.71.033625}, expansion dynamics \cite{PhysLettA.327.397, 
PhysRevA.73.065601, salasnich2007mean}, solitons \cite{EuroPhysLett.79.50003,
PhysRevA.76.043626}, Josephson oscillations \cite{PhysRevA.77.043609}, 
shock wave generation\cite{salasnich2011supersonic,ancilotto2012shock}, etc. 
Recently, NLSE has also been used to investigate the soliton mediated decay 
of shock waves into vortex dipoles in the oblate BCS superfluids 
\cite{gautam2012generation}. In addition to all these, the Bose-Fermi 
superfluid mixtures can be described by the Gross-Pitaevskii equation (GPE) for 
the bosons, coupled to the NLSE for the order parameter of the Fermi superfluid \cite{PhysRevA.76.043626,PhysRevA.78.043616}.
Such coupled equations have been used to study gap solitons 
\cite{PhysRevA.76.043626}, mixing-demixing transition \cite{PhysRevA.76.023612},
the spontaneous symmetry breaking of the Bose-Fermi mixture in double-well potentials \cite{PhysRevA.81.053630},
localization of a Bose-Fermi mixture in a bichromatic optical lattice \cite{PhysRevA.84.023632}, etc. 
The NLSE with quintic repulsive nonlinearity has also been derived to 
describe one dimensional 
Tonks-Girardeau (TG) gas \cite{PhysRevLett.85.1146}, and was used to study 
the formation of shocks and their dynamics \cite{0953-4075-37-5-L01}. 
The same equation, with an additional nonlocal cubic term accounting for 
the dipole-dipole attraction, has been used to study existence, stability 
and dynamics of bright solitons in dipolar one dimensional TG gas 
\cite{0953-4075-42-17-175302}.

In the present work, we theoretically study
the CI in strongly interacting Fermi superfluids using NLSE. The paper is
organized as follows. In Sec. \ref{II}, we describe the NLSE used to study 
the UFG. In Sec. \ref{II}, we study the dynamics of a single vortex in
cigar shaped UFG. In Sec. \ref{IV}, we study the CI in a pair of vortices which
are generated in the UFG by imprinting the appropriate phase. In the
following Sec. \ref{V}, we investigate the generation of vortex-antivortex
pair by a Gaussian obstacle potential, which on subsequent evolution decays
via CI into vortex rings and sound. We conclude by presenting the summary
of the results in the last section.

%%%%%%%%%%%%%%%%%%%%%%%%%%%%%%%%%%%%%%%%%%%%%%%%%%%%%%%%%%%%%%%%%%%%%%%%%%%%%%
%%%%%%%%%%%%            Non-linear Schr\"odinger equation for      %%%%%%%%%%%
%%%%%%%%%%%%                     unitary Fermi gas                 %%%%%%%%%%%
%%%%%%%%%%%%%%%%%%%%%%%%%%%%%%%%%%%%%%%%%%%%%%%%%%%%%%%%%%%%%%%%%%%%%%%%%%%%%%

\section{Non-linear Schr\"odinger equation for unitary Fermi gas}
\label{II}
The order parameter or the condensate wavefunction for the superfluid 
two component fermionic system, excluding the normalization factor, is
\begin{equation}
\psi (\mathbf r, t)  =  \langle\hat\psi_\downarrow(\mathbf r, t)
                                \hat\psi_\uparrow(\mathbf r, t)\rangle,
\end{equation}
where $\psi_\uparrow(\mathbf r, t)$ and $\psi_\downarrow(\mathbf r, t)$ are 
the fermionic annihilation field operators for $\uparrow$ and $\downarrow$ 
spin states respectively. The order parameter can be interpreted as the 
wavefunction of the macroscopically occupied two particle state. The order
parameter is normalized to total number of condensed Cooper pairs 
 \cite{RevModPhys.80.1215},\cite{leggett2006,PhysRevA.72.023621,
PhysRevA.76.015601}. As mentioned earlier, the $s$-wave scattering length 
between two component spin states $a\rightarrow -\infty$ in UFG. The 
Lagrangian of the UFG with energy $E$ and consisting of equal number of two 
spin states at $T=0$ K is \cite{PhysLettA.327.397},\cite{PhysRevA.73.065601},
\cite{salasnich2007mean}, \cite{PhysRevA.76.043626}, 
\cite{PhysRevA.78.043616}, \cite{LaserPhys.4.642}
\begin{equation}
 L = \int d\mathbf r\frac{i\hbar}{2}\left(\psi^*\frac{\partial\psi}
         {\partial t} - \psi\frac{\partial \psi^*}{\partial t }\right)-E.
\end{equation}
%Here $\psi(\mathbf r,t)$ is complex order parameter for the UFG, 
%$m$ is the atomic mass of the fermionic species, $m_p=2m$ is the mass of the 
%fermionic pair, and $V(\mathbf r)$ is the trapping potential. 
The energy of the UFG is given by
\begin{eqnarray}
E & = &\int\left[\frac{\hbar^2}{4 m} |\nabla \psi|^2+ \xi\frac{3\hbar^2}{5m} 
    (3\pi^2)^{2/3}|\psi|^{10/3}\right.\nonumber\\  
  &   &  +  V(\mathbf r,t)|\psi|^2\bigg]d\mathbf r,
\end{eqnarray}
where $m$ is the atomic mass of the fermionic species, $\xi = 0.44$  \cite{PhysRevA.78.043616}, and
$V(\mathbf r,t)$ is the trapping potential.
Using the action principle
\begin{equation}
\delta\int_{t_1}^{t_2}Ldt = 0,
\end{equation}
the time dependent NLSE describing the
UFG at zero temperature is
\begin{eqnarray}
 \Bigg[-\frac{\hbar^2}{4 m} \nabla^2+ \xi\frac{2\hbar^2}{m_p} (3\pi^2)^{2/3}
 |\psi(\mathbf r,t)|^{4/3}   &+&  V(\mathbf r,t)\Bigg]\psi(\mathbf r,t)
  \nonumber\\ 
    &=&  i\hbar\frac{\partial\psi(\mathbf r,t)}{\partial t},
\label{nlse}
\end{eqnarray}
where $m_p=2m$ is the mass of the fermionic pair.
The order parameter $\psi$ satisfies the normalization condition
\begin{equation}
\int |\psi(\mathbf r,t)|^2d\mathbf r = N.
\end{equation}
In the present work, we consider cigar-shaped trapping potential
\begin{equation}
 V(\mathbf r,t) = \frac{m_p\omega^2}{2}(x^2+y^2+\alpha^2 z^2),
\end{equation}
where $\omega$ is the radial trapping frequency, $\alpha<1$ is the ratio of
axial to radial trapping frequency. We consider UFG of $^{40}$K with 
$N = 3500$, $\omega = 100$Hz, and $\alpha = 0.2$. In 
Refs.~\cite{JPhysB.42.215306,NewJPhys.11.023011}, effective one dimensional 
(1D) NLSE for the cigar-shaped Fermi superfluids in the unitary limit has been
proposed. The 1D NLSE provides a good approximation to the real three 
dimensional (3D) systems, provided the dynamics along radial direction is 
frozen. The CI instability being a three dimensional instability can not be 
studied using quasi-1D equation. Hence, in the present work, we consider the 
three dimensional equation and solve it numerically using the split time step 
Crank-Nicolson method \cite{CompPhysComm.180.1888}. In this method, 
the NLSE is discretized in space and time using the finite difference scheme.
The time iteration is performed by splitting the Hamiltonian into two parts, 
one containing the spatial derivative part and the other containing the rest of 
the terms. The error involved in the splitting of the the Hamiltonian is 
proportional to the square of the time step. 

We can rewrite the 
Eq. (\ref{nlse}) in scaled units using the transformations
\begin{eqnarray}
\mathbf r  =  \mathbf r' a_{\rm osc},~t = t' \omega^{-1},\nonumber\\ 
\psi(\mathbf r, t) = \frac{\sqrt{N}\phi(\mathbf r', t')}{a_{\rm osc}^{3/2}}, 
\end{eqnarray}
where the primed quantities are in scaled units and $a_{\rm osc} = 
\sqrt{\hbar/(m_p \omega)}$ is the oscillator length. After dropping the
primes, the scaled NLSE describing the superfluid UFG is
\begin{eqnarray}
 \left[-\frac{\nabla^2}{2} + 2 \xi(3\pi^2 N)^{2/3}
 |\phi|^{4/3}+V(\mathbf r,t)\right]\phi(\mathbf r,t)\nonumber\\ 
 = i \frac{\partial\phi(\mathbf r,t)}{\partial t},
\label{nlse_scaled}
\end{eqnarray}
This equation is used to study the dynamics of UFG in the present work.
%We use split time step Crank-Nicolson method to solve Eq.~(\ref{nlse_scaled})
%\cite{muruganandam2009fortran}. 
We use $\Delta x = \Delta y =\Delta z = 0.15$ and $\Delta t = 0.001$
as the spatial and time step sizes. The number of time iterations used in 
the present work is $55000$. In terms of unscaled units, the spatial and time 
step sizes are $0.169~\mu $m and $1.59\times10^{-3}$~ms respectively. 
%%%%%%%%%%%%%%%%%%%%%%%%%%%%%%%%%%%%%%%%%%%%%%%%%%%%%%%%%%%%%%%%%%%%%%%%%%%%%%%
%%%%%%%%%%%%          CI in vortex antivortex pair generated by    %%%%%%%%%%%%
%%%%%%%%%%%%                   phase-imprinting                    %%%%%%%%%%%%
%%%%%%%%%%%%%%%%%%%%%%%%%%%%%%%%%%%%%%%%%%%%%%%%%%%%%%%%%%%%%%%%%%%%%%%%%%%%%%%

\section{Dynamics of a single vortex line}
\label{III}
In case of BECs, the dynamics of single vortex line in prolate condensates
was studied experimentally in Ref. \cite{rosenbusch2002dynamics}. It was 
observed that vortex line in most cases was bent as was predicted in earlier 
theoretical studies \cite{PhysRevA.63.041603,PhysRevA.64.053611,
PhysRevA.64.043611,PhysRevA.66.023611,PhysRevA.62.063617,PhysRevLett.86.564}. 
It was shown by Svidzinsky et al. \cite{PhysRevA.62.063617} that a straight 
vortex line is stationary only if it is oriented along $z$-axis. The stability
of the vortex line in BECs can be inferred from the normal modes, which can be 
obtained by solving the equation obtained after linearizing the 
Gross-Pitaevskii equation about the stationary configuration. In oblate 
condensates with a vortex line near $z$-axis there is one normal mode with 
negative frequency and it gives the precession frequency of the vortex in 
trap. On the other hand, the appearance of more than one normal modes with 
negative frequency in cigar shaped condensates, makes the vortex line 
increasingly susceptible to bending\cite{PhysRevA.62.063617}. In case of 
cigar-shaped UFG, a similar instability against bending can aid the 
reconnection between the vortex and the antivortex line. Hence, before 
studying the dynamics of vortex-antivortex pair, it can be quite instructive 
to study the dynamics of a single vortex line in prolate UFG.

\begin{figure}
\includegraphics[trim = 0mm 0mm 0mm 0mm,clip, angle=0,width=3.4cm]
                 {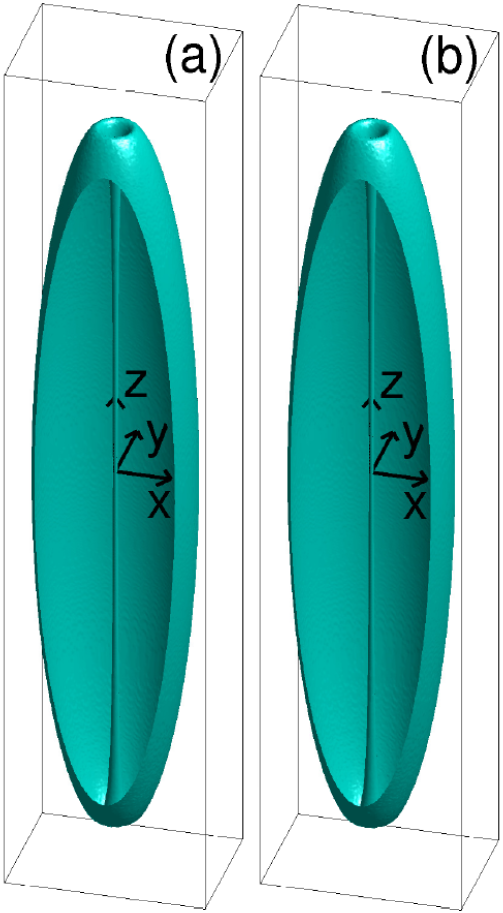}
\caption{The time evolution of the UFG with a single axial vortex line.
The images are the isosurfaces corresponding to $|\phi(\mathbf r)| = 
0.012~a_{\rm osc}^{-3/2}$. Images with the labels (a) and (b) are the 
isosurfaces after $0$ ms and $79.6$ ms  of time evolution respectively.
The origin of the coordinate system is located at the center of the box in 
each image. The dimensions of each box are 
$12.15~a_{\rm osc}\times8.55~a_{\rm osc}\times54.15~a_{\rm osc}$} 
\label{figure1}
\end{figure}

\begin{figure}
\includegraphics[trim = 0mm 0mm 0mm 0mm,clip, angle=0,width=8.4cm]
                 {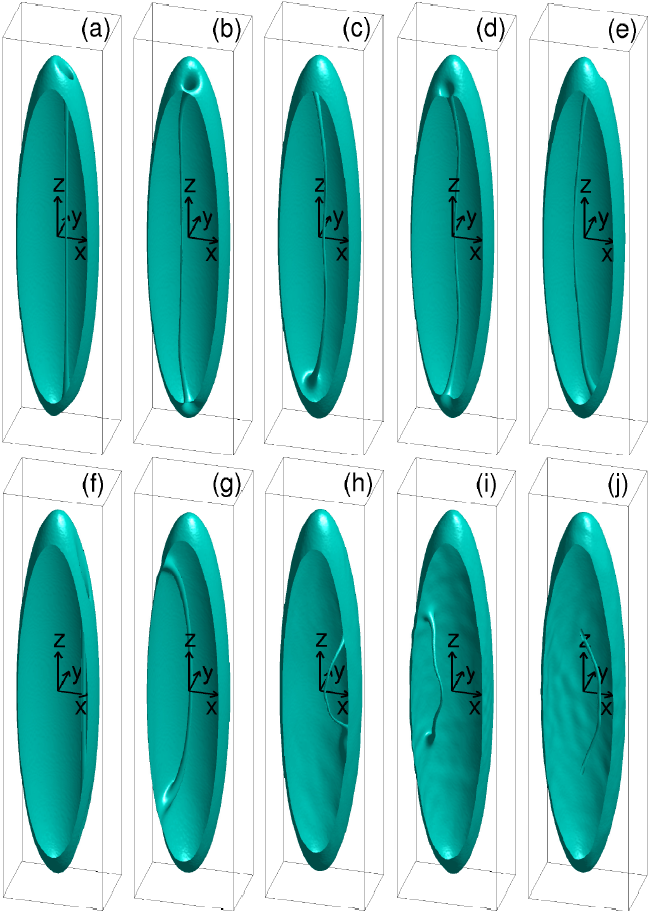}
\caption{Upper panel shows the dynamics of the UFG with an off-axis vortex 
line initially located at $x=1.0~a_{\rm osc}$, whereas the lower panel shows 
the dynamics with an off-axis vortex line initially located at 
$x=3.0~a_{\rm osc}$. The images are the isosurfaces corresponding to 
$|\phi(\mathbf r)| = 0.012~a_{\rm osc}^{-3/2}$. In upper panel, images with 
the labels (a) and (e) are the isosurfaces after $0$ ms, $23.8$ ms, $47.7$ ms, 
$63.7$ ms, and $79.6$ ms  of time evolution respectively. In the lower panel,
images with the labels (f) and (j) are the isosurfaces after $0$ ms, 
$12.7$ ms, $47.7$ ms, $63.7$ ms, and $82.8$ ms respectively. The origin of 
the coordinate system is located at the center of the box in each image. The 
dimensions of each box are 
$12.15~a_{\rm osc}\times8.55~a_{\rm osc}\times54.15~a_{\rm osc}$}
\label{figure2}
\end{figure}

We numerically generate the vortex line by phase-imprinting method, in which 
the order parameter is evolved in imaginary time under the constraint that it 
has the phase consistent with the presence of the vortex line at all times. 
To do so, while  evolving the NLSE in imaginary time $\tau = i t$, the order 
parameter $\phi(\mathbf r,\tau)$  is redefined as
\begin{eqnarray}
\phi(\mathbf r,\tau+\delta \tau) & = &\phi(\mathbf r,\tau)\exp
  \left[i\left(\tan^{-1}\frac{y-y_0}{x-x_0}\right)
  \right]
\label{phase-imprinting}
\end{eqnarray}
after each iteration in imaginary time with the temporal step size of
$\delta\tau$. Here, $(x_0,y_0)$ is the point of intersection of the vortex 
line with $xy$ plane, and $\tan^{-1}[(y-y_0)/(x-x_0)]$ is the azimuthal angle 
measured with the origin shifted to $(x_0,y_0,0)$. This redefinition of the 
order parameter ensures that the solution of the NLSE always has vortex line 
passing through $(x_0,y_0,0)$ and oriented parallel to $z$-axis. The solution
obtained by this method is then evolved in real time to study the dynamics
of the system. We first consider the dynamics of the UFG with an axial vortex 
line. In this case, the vortex line remains straight, and hence is stable 
against the bending instability as is shown in Fig.~\ref{figure1}. This is not 
the case for an off-axis vortex line, which tends to bend as it traverses 
across the condensate. This is evident from Fig.~\ref{figure2}, where the 
upper and lower panel show the dynamics of the vortex line initially located 
at $x=1.0~a_{\rm osc}$ and $x=3.0~a_{\rm osc}$ respectively. Hence, like in BECs,
an off-axis vortex line is unstable against bending in cigar-shaped UFG.
 
 
%%%%%%%%%%%%%%%%%%%%%%%%%%%%%%%%%%%%%%%%%%%%%%%%%%%%%%%%%%%%%%%%%%%%%%%%%%%%%%%
%%%%%%%%%%%%          CI in vortex antivortex pair generated by    %%%%%%%%%%%%
%%%%%%%%%%%%                   phase-imprinting                    %%%%%%%%%%%%
%%%%%%%%%%%%%%%%%%%%%%%%%%%%%%%%%%%%%%%%%%%%%%%%%%%%%%%%%%%%%%%%%%%%%%%%%%%%%%%

\section{CI in vortex-antivortex pair generated by 
phase-imprinting}
\label{IV}
\begin{figure}
\includegraphics[trim = 0mm 0mm 0mm 0mm,clip, angle=0,width=8cm]
                 {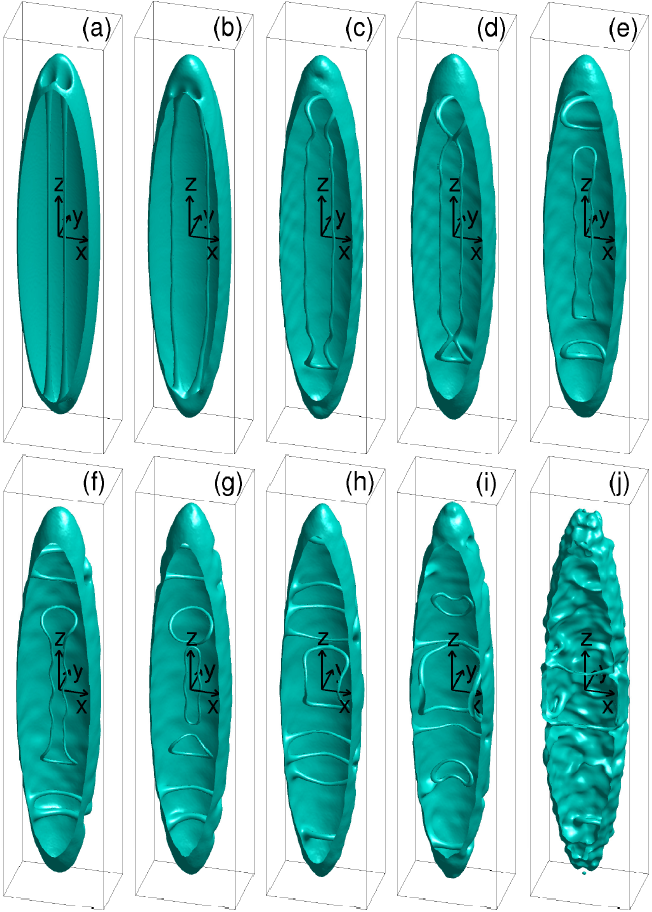}
\caption{(Color online) The time evolution of the cigar-shaped UFG with 
vortex-antivortex imprinted in it at $t = 0$ ms. The images are isosurfaces 
corresponding to $|\phi(\mathbf r)| = 0.012~a_{\rm osc}^{-3/2}$. In the upper 
panel, images with labels (a) to (e) are the isosurfaces after $0$ ms, 
$20.7$ ms, $28.6$ ms, $31.8$ ms, and $35.0$ ms of time evolution respectively. 
In the lower panel, images with labels (f) to (j) are the isosurfaces after 
$38.2$ ms, $41.4$ ms, $47.7$ ms, $54.1$ ms, and $79.6$ ms of time evolution 
respectively. The origin of the coordinate system is located at the center of 
the box in each image. The dimensions of
each box are $12.15~a_{\rm osc}\times8.55~a_{\rm osc}\times54.15~a_{\rm osc}$. 
}
\label{fig1}
\end{figure}

We can extend the phase-imprinting method, described in the previous section,
to the vortex-antivortex pair. In order to numerically imprint a 
vortex-antivortex pair, phase contribution from both the vortex and 
antivortex has to be imprinted on the order parameter after each iteration in 
imaginary time.
%We numerically generate the vortex-antivortex pair by phase-imprinting method.
%In this method, while evolving the NLSE in imaginary time $\tau = i t$, the 
This can be ensured by redefining order parameter $\phi(\mathbf r,\tau)$ as
%order parameter $\phi(\mathbf r,\tau)$  is redefined as
\begin{eqnarray}
\phi(\mathbf r,\tau+\delta \tau) & = &\phi(\mathbf r,\tau)\exp
   \left[i\left(\tan^{-1}\frac{y-y_1}{x-x_1}\right)\right]\nonumber\\
& &\times\exp\left[-i\left(\tan^{-1}\frac{y-y_2}{x-x_2}\right)\right]
   \nonumber\\
                                 & = &\phi(\mathbf r,\tau)\exp
  \left[i\left(\tan^{-1}\frac{y-y_1}{x-x_1}\right.\right.\nonumber\\
         & &\left.\left.-\tan^{-1}\frac{y-y_2}{x-x_2}\right)
  \right]
\end{eqnarray}
after each iteration in imaginary time with the temporal step size of 
$\delta\tau$. Here  $\tan^{-1}[(y-y_1)/(x-x_1)]$ and 
$\tan^{-1}[(y-y_2)/(x-x_2)]$ are the azimuthal angles measured with the origin 
shifted to $(x_1,y_1,0)$ and $(x_2,y_2,0)$ respectively. The converged 
solution obtained by this method will have a vortex
and antivortex at $(x_1,y_1)$ and $(x_2,y_2)$ respectively. This solution is 
then evolved in real time to study the CI. As a case study, we imprint the 
vortex and antivortex at $(-a_{\rm osc},-a_{\rm osc})$ and $(a_{\rm osc},
-a_{\rm osc})$ respectively. The stationary solution obtained by imaginary 
time propagation is shown in Fig. \ref{fig1}(a); here the distance between the
two parallel vortex lines is equal to $2a_{\rm osc}$, and the origin of the 
coordinate system is located at the center of the box.
%\begin{figure}[ht]
The solution is then evolved in the real time using Eq.~(\ref{nlse_scaled}). 
During the initial stages of the evolution vortex-antivortex follow the 
trajectories which is reminiscent of the trajectory followed by a vortex 
dipole in the pancake-shaped traps \cite{PhysRevLett.104.160401}. To be 
precise, the vortex moves in the region to the left of $x=0$ plane, while the 
antivortex moves in the region to the right of  $x=0$ plane. In order to 
understand it, the vortex-antivortex pair at $t=0$ ms can be considered to be 
equivalent to a highly stretched vortex ring with center at $(0,-a_{\rm osc})$.
The motion of this vortex ring will be due to the contribution from two 
velocity fields: a) the self induced velocity which will propel the ring in 
positive $y$ direction and b) the precession induced by the inhomogeneity of 
the condensate. Initially the self induced velocity dominates and the ring 
travels forward, later on the precession effects dominate causing the ring to 
move backward. The precession will result in the expansion of the ring along 
radial direction for $y>0$ and contraction for $y<0$. After the ring returns 
close to its initial position, each element of the ring (vortex and antivortex)
has developed a curvature due to aforementioned expansion as is shown in 
Fig. \ref{fig1}(b). As the pair moves along positive $y$ direction, the 
distance between the vortices at their edges decreases further causing the 
first reconnection event. This results in an actual vortex ring as is shown in
Fig. \ref{fig1}(c). The first reconnection event is aided by bending instability
of the off-center vortices. Due to this bending instability, vortex and antivortex
bend in opposite direction resulting in the decreased separation between them at
the edges [see Fig. \ref{fig1}(b)]. The reconnection event is also accompanied by the emission
of the sound as is evident from ripples on the isosurface in 
Fig. \ref{fig1}(c). Also, in the presence of trapping potential, parts of vortex 
ring in higher density region move slower than those in lower density regions 
\cite{PhysRevA.62.063617}. This leads to the differential velocity field
experienced by the ring. The differential velocity field causes the ring
to bend at the ends [Fig. \ref{fig1}(c)]; this in turn generates the Kelvin 
waves which aid the growth of CI as is shown in Figs. \ref{fig1}(c) and (d). 
For a sufficiently large amplitude of these Kelvin waves, the second 
reconnection event takes place at $31.8$ ms [Fig. \ref{fig1}(d)].
This results in the generation of two daughter vortices and one parent vortex 
[Fig. \ref{fig1}(e)]. Reconnection events are typically accompanied by the 
generation sound waves and further generation of Kelvin waves. With further 
evolution, the parent ring undergoes the same bending at the edges 
[Fig. \ref{fig1}(f)], followed by the generation of second set of daughter 
rings [Fig. \ref{fig1}(g)]. As more time elapses, more reconnection events and
the penetration of the vortex rings in the UFG's surface take place 
[Figs. \ref{fig1}(h-j)]. At the end of the time evolution, the initial 
vortex-antivortex pair has mostly decayed into sound waves due to CI as is 
shown in Fig. \ref{fig1}(j).

%%%%%%%%%%%%%%%%%%%%%%%%%%%%%%%%%%%%%%%%%%%%%%%%%%%%%%%%%%%%%%%%%%%%%%%%%%%%%%%
%%%%%%%%%%%%          CI for vortex-antivortex pair generated by   %%%%%%%%%%%%
%%%%%%%%%%%%                  a moving obstacle                    %%%%%%%%%%%%
%%%%%%%%%%%%%%%%%%%%%%%%%%%%%%%%%%%%%%%%%%%%%%%%%%%%%%%%%%%%%%%%%%%%%%%%%%%%%%%
 
\section{CI in vortex-antivortex pair generated by a moving 
         obstacle potential}
\label{V}
\begin{figure}
\includegraphics[trim = 0mm 0mm 0mm 0mm,clip, angle=0,width=8cm]
                 {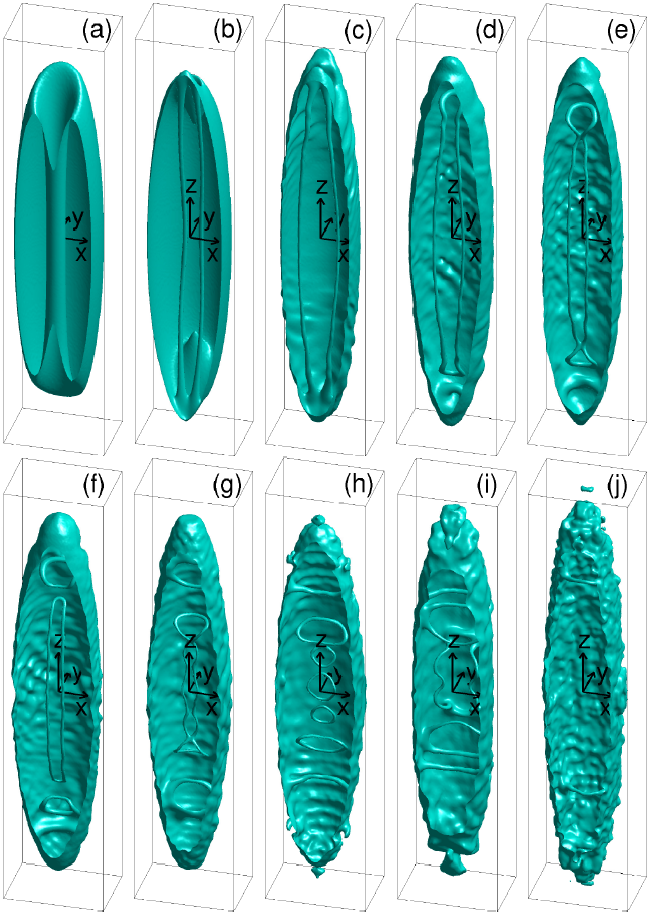}
\caption{(Color online) Evolution of the UFG when the obstacle initially
located at $(0,-a_{\rm osc})$ is moved with the speed of $550~\mu$m/s.
The images are the isosurfaces corresponding to
$|\phi(\mathbf r)| = 0.012~a_{\rm osc}^{-3/2}$. In the upper panel, images
with the labels (a) to (e) are the isosurfaces after $0$ ms, $6.4$ ms, 
$20.7$ ms, $25.5$ ms, and $28.6$ ms of time evolution respectively. 
In the lower panel, 
images with the labels (f) to (j) are the isosurfaces after $30.2$ ms, $35.0$ ms,
$38.2$ ms, $43.0$ ms, and $82.8$ ms of time evolution respectively. The origin of 
the coordinate system is located at the center of the box in each image.
The dimensions of
each box are $12.15~a_{\rm osc}\times8.55~a_{\rm osc}\times54.15~a_{\rm osc}$.}
\label{fig2}
\end{figure}

 Experimentally vortex-antivortex pairs have been created in the BEC by moving
the condensate across a repulsive Gaussian obstacle 
\cite{PhysRevLett.104.160401}. The generation of vortex rings by moving a 
spherical obstacle across a UFG has also been studied theoretically 
\cite{bulgac2011real}. The method of generating vortex antivortex pairs by a 
moving repulsive potential is also a good prospect to experimentally observe 
CI in BECs \cite{PhysRevA.84.021603}. Taking all these into account, we also 
study the evolution of UFG under the influence of a moving obstacle potential.
We employ an obstacle potential
\begin{equation}
V_{\rm obs}(t) = V_0(t) \frac{w_0^2}{w(z)^2}
                 \exp\left\{-2\frac{x^2+[y-y_0(t)]^2}{w(z)^2}\right\}, 
\end{equation}
created by Gaussian laser beam. Here $w_0$ is the waist size, 
$w_z = w_0\sqrt{1-z^2/z_R^2}$, where  $z_R$ is the Rayleigh range, is beam 
width or spot size, and $V_0(t)$ is the maximum potential created by the beam 
at the center of beam waist which lies at $(0,y_0)$. We use the same obstacle
potential parameters as used in Ref.~\cite{PhysRevA.84.021603} : 
$V_0 = 25~\hbar\omega$, $z_R = 22~\mu$m, and $w_0 = 2.2~\mu$m. The stationary 
state isosurface with the obstacle potential located at $(0,-a_{\rm osc})$ is 
shown in Fig. \ref{fig2}(a); the figure clearly shows the depletion of UFG in 
the region of the obstacle potential. We move the obstacle with the speed of 
$500~\mu$m/s along $y$-axis. This speed is a fraction of the speed  of sound 
in UFG at the trap center. We estimate the speed of sound at the trap center 
by using the expression for speed of sound in a homogeneous UFG, i.e., 
$c = \sqrt{\xi}v_{F}/\sqrt{3}$, where $v_F = 2(3\pi^2N|\phi|^2)^{1/3}$ is the 
Fermi velocity for ideal Fermi gas \cite{RevModPhys.80.1215}. Using this 
expression, the speed of the sound at the trap center is $\sim 1800~\mu$m/s.
While moving the obstacle its amplitude is linearly ramped down. The rate
of ramping down is such that the obstacle potential becomes zero at 
$(0,4~a_{\rm osc})$. 
 
%\begin{figure}[ht]
As the obstacle moves along positive $y$ direction, it
creates a vortex antivortex pair in its wake [Fig. \ref{fig2}(b)]. This pair
after being reflected from the opposite surface of the UFG returns closer 
to its initial location [Fig. \ref{fig2}(c)]. This is followed by the first
[Fig. \ref{fig2}(d)] and second reconnection events [Fig. \ref{fig2}(e)]
With further time evolution, CI leads to more reconnection
events, penetration of the vortices in the UFG's surface, and generation of
sound waves [Figs. \ref{fig2}(f-j)] as happened in the case of phase-imprinted
vortex-antivortex pair. At the end of the evolution period,
only sound waves are discernible in the system [Fig. \ref{fig2}(j)].   

%%%%%%%%%%%%%%%%%%%%%%%%%%%%%%%%%%%%%%%%%%%%%%%%%%%%%%%%%%%%%%%%%%%%%%%%%%%%%%%
%%%%%%%%%%%%             Summary of results                        %%%%%%%%%%%%
%%%%%%%%%%%%%%%%%%%%%%%%%%%%%%%%%%%%%%%%%%%%%%%%%%%%%%%%%%%%%%%%%%%%%%%%%%%%%%%

\section{Summary and conclusions}
We have numerically investigated the CI in the cigar-shaped superfluid Fermi
gas in the unitary limit. We use NLSE based on the time dependent density 
functional theory to study CI. We numerically solve this equation using split
time step Crank-Nicolson method to study the dynamics UFG with a single vortex
line as well as a vortex-antivortex pair. In order to generate the
vortex-antivortex pair, we employ two methods: (a) phase-imprinting and (b) 
moving a Gaussian obstacle potential across the system. We have shown that a 
single vortex line in cigar shaped UFG is stable only if it is oriented along 
axial direction. On the other hand, an off-axis vortex line tends to bend as it 
traverses across the condensate. We have shown that a vortex-antivortex pair
oriented along the axial direction in a cigar-shaped UFG is susceptible to
the CI mechanism. We observe that frequent vortex reconnections, inhomogeneity 
generated Kelvin waves, and sound generation characterize the CI in an 
inhomogeneous UFG like in a BEC. We have shown, using the experimentally realizable 
parameters for trapping potential and number of atoms, that it should
be possible to observe the signatures of CI in a vortex-antivortex pair generated
by moving a blue detuned laser beam across the UFG.      
 
%%%%%%%%%%%%%%%%%%%%%%%%%%%%%%%%%%%%%%%%%%%%%%%%%%%%%%%%%%%%%%%%%%%%%%%%%%%%%%%
%%%%%%%%%%%%             Acknowledgements                          %%%%%%%%%%%%
%%%%%%%%%%%%%%%%%%%%%%%%%%%%%%%%%%%%%%%%%%%%%%%%%%%%%%%%%%%%%%%%%%%%%%%%%%%%%%%

\begin{acknowledgements}
We thank Arko Roy, S. Chattopadhyay, Vivek Vyas and Dilip Angom for very 
useful discussions. The numerical computations reported in the paper were 
done on the 3 TFLOPs cluster at PRL. 
\end{acknowledgements}

%%%%%%%%%%%%%%%%%%%%%%%%%%%%%%%%%%%%%%%%%%%%%%%%%%%%%%%%%%%%%%%%%%%%%%%%%%%%%%%
%%%%                        Bibliography                                  %%%%%
%%%%%%%%%%%%%%%%%%%%%%%%%%%%%%%%%%%%%%%%%%%%%%%%%%%%%%%%%%%%%%%%%%%%%%%%%%%%%%%
\bibliography{references.bib} 

%Merlin.mbs v4.21 2009-07-09.
\providecommand{\noopsort}[1]{}\providecommand{\singleletter}[1]{#1}%
\begin{thebibliography}{10}%
\makeatletter
\providecommand \@ifxundefined [1]{%
 \ifx #1\undefined \expandafter \@firstoftwo
 \else \expandafter \@secondoftwo
\fi
}%
\providecommand \@ifnum [1]{%
 \ifnum #1\expandafter \@firstoftwo
 \else \expandafter \@secondoftwo
\fi
}%
\providecommand \enquote [1]{``#1''}%
\providecommand \bibnamefont  [1]{#1}%
\providecommand \bibfnamefont [1]{#1}%
\providecommand \citenamefont [1]{#1}%
\providecommand\href[0]{\@sanitize\@href}%
\providecommand\@href[1]{\endgroup\@@startlink{#1}\endgroup\@@href}%
\providecommand\@@href[1]{#1\@@endlink}%
\providecommand \@sanitize [0]{\begingroup\catcode`\&12\catcode`\#12\relax}%
\@ifxundefined \pdfoutput {\@firstoftwo}{%
 \@ifnum{\z@=\pdfoutput}{\@firstoftwo}{\@secondoftwo}%
}{%
 \providecommand\@@startlink[1]{\leavevmode\special{html:<a href="#1">}}%
 \providecommand\@@endlink[0]{\special{html:</a>}}%
}{%
 \providecommand\@@startlink[1]{%
  \leavevmode
  \pdfstartlink
   attr{/Border[0 0 1 ]/H/I/C[0 1 1]}%
   user{/Subtype/Link/A<</Type/Action/S/URI/URI(#1)>>}%
  \relax
 }%
 \providecommand\@@endlink[0]{\pdfendlink}%
}%
\providecommand \url  [0]{\begingroup\@sanitize \@url }%
\providecommand \@url [1]{\endgroup\@href {#1}{\urlprefix}}%
\providecommand \urlprefix [0]{URL }%
\providecommand \Eprint[0]{\href }%
\@ifxundefined \urlstyle {%
  \providecommand \doi [1]{doi:\discretionary{}{}{}#1}%
}{%
  \providecommand \doi [0]{doi:\discretionary{}{}{}\begingroup
  \urlstyle{rm}\Url }%
}%
\providecommand \doibase [0]{http://dx.doi.org/}%
\providecommand \Doi[1]{\href{\doibase#1}}%
\providecommand \bibAnnote [3]{%
  \BibitemShut{#1}%
  \begin{quotation}\noindent
    \textsc{Key:}\ #2\\\textsc{Annotation:}\ #3%
  \end{quotation}%
}%
\providecommand \bibAnnoteFile [2]{%
  \IfFileExists{#2}{\bibAnnote {#1} {#2} {\input{#2}}}{}%
}%
\providecommand \typeout [0]{\immediate \write \m@ne }%
\providecommand \selectlanguage [0]{\@gobble}%
\providecommand \bibinfo [0]{\@secondoftwo}%
\providecommand \bibfield [0]{\@secondoftwo}%
\providecommand \translation [1]{[#1]}%
\providecommand \BibitemOpen[0]{}%
\providecommand \bibitemStop [0]{}%
\providecommand \bibitemNoStop [0]{.\EOS\space}%
\providecommand \EOS [0]{\spacefactor3000\relax}%
\providecommand \BibitemShut [1]{\csname bibitem#1\endcsname}%
%</preamble>
\bibitem{crow.8.2172}%
  \BibitemOpen
  \bibfield{author}{%
  \bibinfo {author} {\bibfnamefont{S.}~\bibnamefont{Crow}},\ }%
  \bibfield{journal}{%
  \bibinfo {journal} {AIAA Journal}\ }%
  \textbf{\bibinfo {volume} {8}},\ \bibinfo {pages} {2172} (\bibinfo {year}
  {1970})%
  \bibAnnoteFile{NoStop}{crow.8.2172}%
\bibitem{Kelvin}%
  \BibitemOpen
  \bibfield{author}{%
  \bibinfo {author} {\bibfnamefont{W.}~\bibnamefont{Thomson (Lord~Kelvin)}},\
  }%
  \bibfield{journal}{%
  \bibinfo {journal} {Philos. Mag.}\ }%
  \textbf{\bibinfo {volume} {10}},\ \bibinfo {pages} {155} (\bibinfo {year}
  {1880})%
  \bibAnnoteFile{NoStop}{Kelvin}%
\bibitem{PhysRevE.51.4479}%
  \BibitemOpen
  \bibfield{author}{%
  \bibinfo {author} {\bibfnamefont{E.~A.}\ \bibnamefont{Kuznetsov}}\ and\
  \bibinfo {author} {\bibfnamefont{J.~J.}\ \bibnamefont{Rasmussen}},\ }%
  \bibfield{journal}{%
  \Doi{10.1103/PhysRevE.51.4479}{\bibinfo {journal} {Phys. Rev. E}}\ }%
  \textbf{\bibinfo {volume} {51}},\ \bibinfo {pages} {4479} (\bibinfo {year}
  {1995})%
  \bibAnnoteFile{NoStop}{PhysRevE.51.4479}%
\bibitem{onsager.6.279}%
  \BibitemOpen
  \bibfield{author}{%
  \bibinfo {author} {\bibfnamefont{L.}~\bibnamefont{Onsager}},\ }%
  \bibfield{journal}{%
  \bibinfo {journal} {Il Nuovo Cimento (1943-1954)}\ }%
  \textbf{\bibinfo {volume} {6}},\ \bibinfo {pages} {279} (\bibinfo {year}
  {1949})%
  \bibAnnoteFile{NoStop}{onsager.6.279}%
\bibitem{feynman1955}%
  \BibitemOpen
  \bibfield{author}{%
  \bibinfo {author} {\bibfnamefont{R.}~\bibnamefont{Feynman}},\ }%
  \emph{\bibinfo {title} {Progress in low temperature physics}},\ Vol.~\bibinfo
  {volume} {1}\ (\bibinfo {publisher} {North Holland},\ \bibinfo {year}
  {1955})%
  \bibAnnoteFile{NoStop}{feynman1955}%
\bibitem{PhysRevLett.83.2498}%
  \BibitemOpen
  \bibfield{author}{%
  \bibinfo {author} {\bibfnamefont{M.~R.}\ \bibnamefont{Matthews}}, \bibinfo
  {author} {\bibfnamefont{B.~P.}\ \bibnamefont{Anderson}}, \bibinfo {author}
  {\bibfnamefont{P.~C.}\ \bibnamefont{Haljan}}, \bibinfo {author}
  {\bibfnamefont{D.~S.}\ \bibnamefont{Hall}}, \bibinfo {author}
  {\bibfnamefont{C.~E.}\ \bibnamefont{Wieman}},\ and\ \bibinfo {author}
  {\bibfnamefont{E.~A.}\ \bibnamefont{Cornell}},\ }%
  \bibfield{journal}{%
  \Doi{10.1103/PhysRevLett.83.2498}{\bibinfo {journal} {Phys. Rev. Lett.}}\ }%
  \textbf{\bibinfo {volume} {83}},\ \bibinfo {pages} {2498} (\bibinfo {year}
  {1999})%
  \bibAnnoteFile{NoStop}{PhysRevLett.83.2498}%
\bibitem{PhysRevLett.84.806}%
  \BibitemOpen
  \bibfield{author}{%
  \bibinfo {author} {\bibfnamefont{K.~W.}\ \bibnamefont{Madison}}, \bibinfo
  {author} {\bibfnamefont{F.}~\bibnamefont{Chevy}}, \bibinfo {author}
  {\bibfnamefont{W.}~\bibnamefont{Wohlleben}},\ and\ \bibinfo {author}
  {\bibfnamefont{J.}~\bibnamefont{Dalibard}},\ }%
  \bibfield{journal}{%
  \Doi{10.1103/PhysRevLett.84.806}{\bibinfo {journal} {Phys. Rev. Lett.}}\ }%
  \textbf{\bibinfo {volume} {84}},\ \bibinfo {pages} {806} (\bibinfo {year}
  {2000})%
  \bibAnnoteFile{NoStop}{PhysRevLett.84.806}%
\bibitem{Science.292.476}%
  \BibitemOpen
  \bibfield{author}{%
  \bibinfo {author} {\bibfnamefont{J.}~\bibnamefont{Abo-Shaeer}}, \bibinfo
  {author} {\bibfnamefont{C.}~\bibnamefont{Raman}}, \bibinfo {author}
  {\bibfnamefont{J.}~\bibnamefont{Vogels}},\ and\ \bibinfo {author}
  {\bibfnamefont{W.}~\bibnamefont{Ketterle}},\ }%
  \bibfield{journal}{%
  \bibinfo {journal} {Science}\ }%
  \textbf{\bibinfo {volume} {292}},\ \bibinfo {pages} {476} (\bibinfo {year}
  {2001})%
  \bibAnnoteFile{NoStop}{Science.292.476}%
\bibitem{Nature.332.1288}%
  \BibitemOpen
  \bibfield{author}{%
  \bibinfo {author} {\bibfnamefont{M.}~\bibnamefont{Zwierlein}}, \bibinfo
  {author} {\bibfnamefont{J.}~\bibnamefont{Abo-Shaeer}}, \bibinfo {author}
  {\bibfnamefont{A.}~\bibnamefont{Schirotzek}}, \bibinfo {author}
  {\bibfnamefont{C.}~\bibnamefont{Schunck}},\ and\ \bibinfo {author}
  {\bibfnamefont{W.}~\bibnamefont{Ketterle}},\ }%
  \bibfield{journal}{%
  \bibinfo {journal} {Nature}\ }%
  \textbf{\bibinfo {volume} {435}},\ \bibinfo {pages} {1047} (\bibinfo {year}
  {2005})%
  \bibAnnoteFile{NoStop}{Nature.332.1288}%
\bibitem{JPhysA.34.10057}%
  \BibitemOpen
  \bibfield{author}{%
  \bibinfo {author} {\bibfnamefont{N.~G.}\ \bibnamefont{Berloff}}\ and\
  \bibinfo {author} {\bibfnamefont{P.~H.}\ \bibnamefont{Roberts}},\ }%
  \bibfield{journal}{%
  \bibinfo {journal} {Journal of Physics A: Mathematical and General}\ }%
  \textbf{\bibinfo {volume} {34}},\ \bibinfo {pages} {10057} (\bibinfo {year}
  {2001})%
  \bibAnnoteFile{NoStop}{JPhysA.34.10057}%
\bibitem{kadomtsev.192.753}%
  \BibitemOpen
  \bibfield{author}{%
  \bibinfo {author} {\bibfnamefont{B.~B.}\ \bibnamefont{Kadomtsev}}\ and\
  \bibinfo {author} {\bibfnamefont{V.~I.}\ \bibnamefont{Petviashvili}},\ }%
  \bibfield{journal}{%
  \bibinfo {journal} {Dokl. Akad. Nauk SSSR}\ }%
  \textbf{\bibinfo {volume} {192}},\ \bibinfo {pages} {753} (\bibinfo {year}
  {1970})%
  \bibAnnoteFile{NoStop}{kadomtsev.192.753}%
\bibitem{PhysRevA.84.021603}%
  \BibitemOpen
  \bibfield{author}{%
  \bibinfo {author} {\bibfnamefont{T.~P.}\ \bibnamefont{Simula}},\ }%
  \bibfield{journal}{%
  \Doi{10.1103/PhysRevA.84.021603}{\bibinfo {journal} {Phys. Rev. A}}\ }%
  \textbf{\bibinfo {volume} {84}},\ \bibinfo {pages} {021603} (\bibinfo {year}
  {2011})%
  \bibAnnoteFile{NoStop}{PhysRevA.84.021603}%
\bibitem{PhysRevLett.86.2926}%
  \BibitemOpen
  \bibfield{author}{%
  \bibinfo {author} {\bibfnamefont{B.~P.}\ \bibnamefont{Anderson}}, \bibinfo
  {author} {\bibfnamefont{P.~C.}\ \bibnamefont{Haljan}}, \bibinfo {author}
  {\bibfnamefont{C.~A.}\ \bibnamefont{Regal}}, \bibinfo {author}
  {\bibfnamefont{D.~L.}\ \bibnamefont{Feder}}, \bibinfo {author}
  {\bibfnamefont{L.~A.}\ \bibnamefont{Collins}}, \bibinfo {author}
  {\bibfnamefont{C.~W.}\ \bibnamefont{Clark}},\ and\ \bibinfo {author}
  {\bibfnamefont{E.~A.}\ \bibnamefont{Cornell}},\ }%
  \bibfield{journal}{%
  \Doi{10.1103/PhysRevLett.86.2926}{\bibinfo {journal} {Phys. Rev. Lett.}}\ }%
  \textbf{\bibinfo {volume} {86}},\ \bibinfo {pages} {2926} (\bibinfo {year}
  {2001})%
  \bibAnnoteFile{NoStop}{PhysRevLett.86.2926}%
\bibitem{Science.293.663}%
  \BibitemOpen
  \bibfield{author}{%
  \bibinfo {author} {\bibfnamefont{Z.}~\bibnamefont{Dutton}}, \bibinfo {author}
  {\bibfnamefont{M.}~\bibnamefont{Budde}}, \bibinfo {author}
  {\bibfnamefont{C.}~\bibnamefont{Slowe}},\ and\ \bibinfo {author}
  {\bibfnamefont{L.}~\bibnamefont{Hau}},\ }%
  \bibfield{journal}{%
  \bibinfo {journal} {Science}\ }%
  \textbf{\bibinfo {volume} {293}},\ \bibinfo {pages} {663} (\bibinfo {year}
  {2001})%
  \bibAnnoteFile{NoStop}{Science.293.663}%
\bibitem{PhysRevLett.104.160401}%
  \BibitemOpen
  \bibfield{author}{%
  \bibinfo {author} {\bibfnamefont{T.~W.}\ \bibnamefont{Neely}}, \bibinfo
  {author} {\bibfnamefont{E.~C.}\ \bibnamefont{Samson}}, \bibinfo {author}
  {\bibfnamefont{A.~S.}\ \bibnamefont{Bradley}}, \bibinfo {author}
  {\bibfnamefont{M.~J.}\ \bibnamefont{Davis}},\ and\ \bibinfo {author}
  {\bibfnamefont{B.~P.}\ \bibnamefont{Anderson}},\ }%
  \bibfield{journal}{%
  \Doi{10.1103/PhysRevLett.104.160401}{\bibinfo {journal} {Phys. Rev. Lett.}}\
  }%
  \textbf{\bibinfo {volume} {104}},\ \bibinfo {pages} {160401} (\bibinfo {year}
  {2010})%
  \bibAnnoteFile{NoStop}{PhysRevLett.104.160401}%
\bibitem{PhysRevLett.90.100403}%
  \BibitemOpen
  \bibfield{author}{%
  \bibinfo {author} {\bibfnamefont{V.}~\bibnamefont{Bretin}}, \bibinfo {author}
  {\bibfnamefont{P.}~\bibnamefont{Rosenbusch}}, \bibinfo {author}
  {\bibfnamefont{F.}~\bibnamefont{Chevy}}, \bibinfo {author}
  {\bibfnamefont{G.~V.}\ \bibnamefont{Shlyapnikov}},\ and\ \bibinfo {author}
  {\bibfnamefont{J.}~\bibnamefont{Dalibard}},\ }%
  \bibfield{journal}{%
  \Doi{10.1103/PhysRevLett.90.100403}{\bibinfo {journal} {Phys. Rev. Lett.}}\
  }%
  \textbf{\bibinfo {volume} {90}},\ \bibinfo {pages} {100403} (\bibinfo {year}
  {2003})%
  \bibAnnoteFile{NoStop}{PhysRevLett.90.100403}%
\bibitem{PhysRevLett.101.020402}%
  \BibitemOpen
  \bibfield{author}{%
  \bibinfo {author} {\bibfnamefont{T.~P.}\ \bibnamefont{Simula}}, \bibinfo
  {author} {\bibfnamefont{T.}~\bibnamefont{Mizushima}},\ and\ \bibinfo {author}
  {\bibfnamefont{K.}~\bibnamefont{Machida}},\ }%
  \bibfield{journal}{%
  \Doi{10.1103/PhysRevLett.101.020402}{\bibinfo {journal} {Phys. Rev. Lett.}}\
  }%
  \textbf{\bibinfo {volume} {101}},\ \bibinfo {pages} {020402} (\bibinfo {year}
  {2008})%
  \bibAnnoteFile{NoStop}{PhysRevLett.101.020402}%
\bibitem{PhysRevA.84.023637}%
  \BibitemOpen
  \bibfield{author}{%
  \bibinfo {author} {\bibfnamefont{S.~J.}\ \bibnamefont{Rooney}}, \bibinfo
  {author} {\bibfnamefont{P.~B.}\ \bibnamefont{Blakie}}, \bibinfo {author}
  {\bibfnamefont{B.~P.}\ \bibnamefont{Anderson}},\ and\ \bibinfo {author}
  {\bibfnamefont{A.~S.}\ \bibnamefont{Bradley}},\ }%
  \bibfield{journal}{%
  \Doi{10.1103/PhysRevA.84.023637}{\bibinfo {journal} {Phys. Rev. A}}\ }%
  \textbf{\bibinfo {volume} {84}},\ \bibinfo {pages} {023637} (\bibinfo {year}
  {2011})%
  \bibAnnoteFile{NoStop}{PhysRevA.84.023637}%
\bibitem{RevModPhys.80.1215}%
  \BibitemOpen
  \bibfield{author}{%
  \bibinfo {author} {\bibfnamefont{S.}~\bibnamefont{Giorgini}}, \bibinfo
  {author} {\bibfnamefont{L.}~\bibnamefont{Pitaevskii}},\ and\ \bibinfo
  {author} {\bibfnamefont{S.}~\bibnamefont{Stringari}},\ }%
  \bibfield{journal}{%
  \bibinfo {journal} {Rev. Mod. Phys.}\ }%
  \textbf{\bibinfo {volume} {80}},\ \bibinfo {pages} {1215} (\bibinfo {year}
  {2008})%
  \bibAnnoteFile{NoStop}{RevModPhys.80.1215}%
\bibitem{bulgac2011real}%
  \BibitemOpen
  \bibfield{author}{%
  \bibinfo {author} {\bibfnamefont{A.}~\bibnamefont{Bulgac}}, \bibinfo {author}
  {\bibfnamefont{Y.}~\bibnamefont{Luo}}, \bibinfo {author}
  {\bibfnamefont{P.}~\bibnamefont{Magierski}}, \bibinfo {author}
  {\bibfnamefont{K.}~\bibnamefont{Roche}},\ and\ \bibinfo {author}
  {\bibfnamefont{Y.}~\bibnamefont{Yu}},\ }%
  \bibfield{journal}{%
  \bibinfo {journal} {Science}\ }%
  \textbf{\bibinfo {volume} {332}},\ \bibinfo {pages} {1288} (\bibinfo {year}
  {2011})%
  \bibAnnoteFile{NoStop}{bulgac2011real}%
\bibitem{PhysLettA.327.397}%
  \BibitemOpen
  \bibfield{author}{%
  \bibinfo {author} {\bibfnamefont{Y.}~\bibnamefont{Kim}}\ and\ \bibinfo
  {author} {\bibfnamefont{A.}~\bibnamefont{Zubarev}},\ }%
  \bibfield{journal}{%
  \bibinfo {journal} {Physics Letters A}\ }%
  \textbf{\bibinfo {volume} {327}},\ \bibinfo {pages} {397} (\bibinfo {year}
  {2004})%
  \bibAnnoteFile{NoStop}{PhysLettA.327.397}%
\bibitem{PhysRevA.70.033612}%
  \BibitemOpen
  \bibfield{author}{%
  \bibinfo {author} {\bibfnamefont{Y.~E.}\ \bibnamefont{Kim}}\ and\ \bibinfo
  {author} {\bibfnamefont{A.~L.}\ \bibnamefont{Zubarev}},\ }%
  \bibfield{journal}{%
  \Doi{10.1103/PhysRevA.70.033612}{\bibinfo {journal} {Phys. Rev. A}}\ }%
  \textbf{\bibinfo {volume} {70}},\ \bibinfo {pages} {033612} (\bibinfo {year}
  {2004})%
  \bibAnnoteFile{NoStop}{PhysRevA.70.033612}%
\bibitem{PhysRevA.71.033625}%
  \BibitemOpen
  \bibfield{author}{%
  \bibinfo {author} {\bibfnamefont{N.}~\bibnamefont{Manini}}\ and\ \bibinfo
  {author} {\bibfnamefont{L.}~\bibnamefont{Salasnich}},\ }%
  \bibfield{journal}{%
  \Doi{10.1103/PhysRevA.71.033625}{\bibinfo {journal} {Phys. Rev. A}}\ }%
  \textbf{\bibinfo {volume} {71}},\ \bibinfo {pages} {033625} (\bibinfo {year}
  {2005})%
  \bibAnnoteFile{NoStop}{PhysRevA.71.033625}%
\bibitem{PhysRevA.73.065601}%
  \BibitemOpen
  \bibfield{author}{%
  \bibinfo {author} {\bibfnamefont{G.}~\bibnamefont{Diana}}, \bibinfo {author}
  {\bibfnamefont{N.}~\bibnamefont{Manini}},\ and\ \bibinfo {author}
  {\bibfnamefont{L.}~\bibnamefont{Salasnich}},\ }%
  \bibfield{journal}{%
  \Doi{10.1103/PhysRevA.73.065601}{\bibinfo {journal} {Phys. Rev. A}}\ }%
  \textbf{\bibinfo {volume} {73}},\ \bibinfo {pages} {065601} (\bibinfo {year}
  {2006})%
  \bibAnnoteFile{NoStop}{PhysRevA.73.065601}%
\bibitem{salasnich2007mean}%
  \BibitemOpen
  \bibfield{author}{%
  \bibinfo {author} {\bibfnamefont{L.}~\bibnamefont{Salasnich}}\ and\ \bibinfo
  {author} {\bibfnamefont{N.}~\bibnamefont{Manini}},\ }%
  \bibfield{journal}{%
  \bibinfo {journal} {Laser physics}\ }%
  \textbf{\bibinfo {volume} {17}},\ \bibinfo {pages} {169} (\bibinfo {year}
  {2007})%
  \bibAnnoteFile{NoStop}{salasnich2007mean}%
\bibitem{EuroPhysLett.79.50003}%
  \BibitemOpen
  \bibfield{author}{%
  \bibinfo {author} {\bibfnamefont{S.}~\bibnamefont{Adhikari}}\ and\ \bibinfo
  {author} {\bibfnamefont{B.}~\bibnamefont{Malomed}},\ }%
  \bibfield{journal}{%
  \bibinfo {journal} {EPL (Europhysics Letters)}\ }%
  \textbf{\bibinfo {volume} {79}},\ \bibinfo {pages} {50003} (\bibinfo {year}
  {2007})%
  \bibAnnoteFile{NoStop}{EuroPhysLett.79.50003}%
\bibitem{PhysRevA.76.043626}%
  \BibitemOpen
  \bibfield{author}{%
  \bibinfo {author} {\bibfnamefont{S.~K.}\ \bibnamefont{Adhikari}}\ and\
  \bibinfo {author} {\bibfnamefont{B.~A.}\ \bibnamefont{Malomed}},\ }%
  \bibfield{journal}{%
  \Doi{10.1103/PhysRevA.76.043626}{\bibinfo {journal} {Phys. Rev. A}}\ }%
  \textbf{\bibinfo {volume} {76}},\ \bibinfo {pages} {043626} (\bibinfo {year}
  {2007})%
  \bibAnnoteFile{NoStop}{PhysRevA.76.043626}%
\bibitem{PhysRevA.77.043609}%
  \BibitemOpen
  \bibfield{author}{%
  \bibinfo {author} {\bibfnamefont{L.}~\bibnamefont{Salasnich}}, \bibinfo
  {author} {\bibfnamefont{N.}~\bibnamefont{Manini}},\ and\ \bibinfo {author}
  {\bibfnamefont{F.}~\bibnamefont{Toigo}},\ }%
  \bibfield{journal}{%
  \Doi{10.1103/PhysRevA.77.043609}{\bibinfo {journal} {Phys. Rev. A}}\ }%
  \textbf{\bibinfo {volume} {77}},\ \bibinfo {pages} {043609} (\bibinfo {year}
  {2008})%
  \bibAnnoteFile{NoStop}{PhysRevA.77.043609}%
\bibitem{PhysRevA.78.053626}%
  \BibitemOpen
  \bibfield{author}{%
  \bibinfo {author} {\bibfnamefont{L.}~\bibnamefont{Salasnich}}\ and\ \bibinfo
  {author} {\bibfnamefont{F.}~\bibnamefont{Toigo}},\ }%
  \bibfield{journal}{%
  \Doi{10.1103/PhysRevA.78.053626}{\bibinfo {journal} {Phys. Rev. A}}\ }%
  \textbf{\bibinfo {volume} {78}},\ \bibinfo {pages} {053626} (\bibinfo {year}
  {2008})%
  \bibAnnoteFile{NoStop}{PhysRevA.78.053626}%
\bibitem{PhysRevA.78.043616}%
  \BibitemOpen
  \bibfield{author}{%
  \bibinfo {author} {\bibfnamefont{S.~K.}\ \bibnamefont{Adhikari}}\ and\
  \bibinfo {author} {\bibfnamefont{L.}~\bibnamefont{Salasnich}},\ }%
  \bibfield{journal}{%
  \Doi{10.1103/PhysRevA.78.043616}{\bibinfo {journal} {Phys. Rev. A}}\ }%
  \textbf{\bibinfo {volume} {78}},\ \bibinfo {pages} {043616} (\bibinfo {year}
  {2008})%
  \bibAnnoteFile{NoStop}{PhysRevA.78.043616}%
\bibitem{PhysRevA.77.045602}%
  \BibitemOpen
  \bibfield{author}{%
  \bibinfo {author} {\bibfnamefont{S.~K.}\ \bibnamefont{Adhikari}},\ }%
  \bibfield{journal}{%
  \Doi{10.1103/PhysRevA.77.045602}{\bibinfo {journal} {Phys. Rev. A}}\ }%
  \textbf{\bibinfo {volume} {77}},\ \bibinfo {pages} {045602} (\bibinfo {year}
  {2008})%
  \bibAnnoteFile{NoStop}{PhysRevA.77.045602}%
\bibitem{PhysRevA.79.023611}%
  \BibitemOpen
  \bibfield{author}{%
  \bibinfo {author} {\bibfnamefont{S.~K.}\ \bibnamefont{Adhikari}},\ }%
  \bibfield{journal}{%
  \Doi{10.1103/PhysRevA.79.023611}{\bibinfo {journal} {Phys. Rev. A}}\ }%
  \textbf{\bibinfo {volume} {79}},\ \bibinfo {pages} {023611} (\bibinfo {year}
  {2009})%
  \bibAnnoteFile{NoStop}{PhysRevA.79.023611}%
\bibitem{JPhysB.42.215306}%
  \BibitemOpen
  \bibfield{author}{%
  \bibinfo {author} {\bibfnamefont{C.}~\bibnamefont{Buitrago}}\ and\ \bibinfo
  {author} {\bibfnamefont{S.}~\bibnamefont{Adhikari}},\ }%
  \bibfield{journal}{%
  \bibinfo {journal} {J Phys. B: At. Mol. Opt. Phys.}\ }%
  \textbf{\bibinfo {volume} {42}},\ \bibinfo {pages} {215306} (\bibinfo {year}
  {2009})%
  \bibAnnoteFile{NoStop}{JPhysB.42.215306}%
\bibitem{NewJPhys.11.023011}%
  \BibitemOpen
  \bibfield{author}{%
  \bibinfo {author} {\bibfnamefont{S.}~\bibnamefont{Adhikari}}\ and\ \bibinfo
  {author} {\bibfnamefont{L.}~\bibnamefont{Salasnich}},\ }%
  \bibfield{journal}{%
  \bibinfo {journal} {New Journal of Physics}\ }%
  \textbf{\bibinfo {volume} {11}},\ \bibinfo {pages} {023011} (\bibinfo {year}
  {2009})%
  \bibAnnoteFile{NoStop}{NewJPhys.11.023011}%
\bibitem{LaserPhys.4.642}%
  \BibitemOpen
  \bibfield{author}{%
  \bibinfo {author} {\bibfnamefont{L.}~\bibnamefont{Salasnich}},\ }%
  \bibfield{journal}{%
  \bibinfo {journal} {Laser physics}\ }%
  \textbf{\bibinfo {volume} {19}},\ \bibinfo {pages} {642} (\bibinfo {year}
  {2009})%
  \bibAnnoteFile{NoStop}{LaserPhys.4.642}%
\bibitem{salasnich2011supersonic}%
  \BibitemOpen
  \bibfield{author}{%
  \bibinfo {author} {\bibfnamefont{L.}~\bibnamefont{Salasnich}},\ }%
  \bibfield{journal}{%
  \bibinfo {journal} {EPL (Europhysics Letters)}\ }%
  \textbf{\bibinfo {volume} {96}},\ \bibinfo {pages} {40007} (\bibinfo {year}
  {2011})%
  \bibAnnoteFile{NoStop}{salasnich2011supersonic}%
\bibitem{ancilotto2012shock}%
  \BibitemOpen
  \bibfield{author}{%
  \bibinfo {author} {\bibfnamefont{F.}~\bibnamefont{Ancilotto}}, \bibinfo
  {author} {\bibfnamefont{L.}~\bibnamefont{Salasnich}},\ and\ \bibinfo {author}
  {\bibfnamefont{F.}~\bibnamefont{Toigo}},\ }%
  \bibfield{journal}{%
  \bibinfo {journal} {Arxiv preprint arXiv:1206.0568}}%
   (\bibinfo {year} {2012})%
  \bibAnnoteFile{NoStop}{ancilotto2012shock}%
\bibitem{gautam2012generation}%
  \BibitemOpen
  \bibfield{author}{%
  \bibinfo {author} {\bibfnamefont{S.}~\bibnamefont{Gautam}},\ }%
  \bibfield{journal}{%
  \bibinfo {journal} {Arxiv preprint arXiv:1205.5670}}%
   (\bibinfo {year} {2012})%
  \bibAnnoteFile{NoStop}{gautam2012generation}%
\bibitem{PhysRevA.76.023612}%
  \BibitemOpen
  \bibfield{author}{%
  \bibinfo {author} {\bibfnamefont{S.~K.}\ \bibnamefont{Adhikari}}\ and\
  \bibinfo {author} {\bibfnamefont{L.}~\bibnamefont{Salasnich}},\ }%
  \bibfield{journal}{%
  \Doi{10.1103/PhysRevA.76.023612}{\bibinfo {journal} {Phys. Rev. A}}\ }%
  \textbf{\bibinfo {volume} {76}},\ \bibinfo {pages} {023612} (\bibinfo {year}
  {2007})%
  \bibAnnoteFile{NoStop}{PhysRevA.76.023612}%
\bibitem{PhysRevA.81.053630}%
  \BibitemOpen
  \bibfield{author}{%
  \bibinfo {author} {\bibfnamefont{S.~K.}\ \bibnamefont{Adhikari}}, \bibinfo
  {author} {\bibfnamefont{B.~A.}\ \bibnamefont{Malomed}}, \bibinfo {author}
  {\bibfnamefont{L.}~\bibnamefont{Salasnich}},\ and\ \bibinfo {author}
  {\bibfnamefont{F.}~\bibnamefont{Toigo}},\ }%
  \bibfield{journal}{%
  \Doi{10.1103/PhysRevA.81.053630}{\bibinfo {journal} {Phys. Rev. A}}\ }%
  \textbf{\bibinfo {volume} {81}},\ \bibinfo {pages} {053630} (\bibinfo {year}
  {2010})%
  \bibAnnoteFile{NoStop}{PhysRevA.81.053630}%
\bibitem{PhysRevA.84.023632}%
  \BibitemOpen
  \bibfield{author}{%
  \bibinfo {author} {\bibfnamefont{Y.}~\bibnamefont{Cheng}}\ and\ \bibinfo
  {author} {\bibfnamefont{S.~K.}\ \bibnamefont{Adhikari}},\ }%
  \bibfield{journal}{%
  \Doi{10.1103/PhysRevA.84.023632}{\bibinfo {journal} {Phys. Rev. A}}\ }%
  \textbf{\bibinfo {volume} {84}},\ \bibinfo {pages} {023632} (\bibinfo {year}
  {2011})%
  \bibAnnoteFile{NoStop}{PhysRevA.84.023632}%
\bibitem{PhysRevLett.85.1146}%
  \BibitemOpen
  \bibfield{author}{%
  \bibinfo {author} {\bibfnamefont{E.~B.}\ \bibnamefont{Kolomeisky}}, \bibinfo
  {author} {\bibfnamefont{T.~J.}\ \bibnamefont{Newman}}, \bibinfo {author}
  {\bibfnamefont{J.~P.}\ \bibnamefont{Straley}},\ and\ \bibinfo {author}
  {\bibfnamefont{X.}~\bibnamefont{Qi}},\ }%
  \bibfield{journal}{%
  \Doi{10.1103/PhysRevLett.85.1146}{\bibinfo {journal} {Phys. Rev. Lett.}}\ }%
  \textbf{\bibinfo {volume} {85}},\ \bibinfo {pages} {1146} (\bibinfo {year}
  {2000})%
  \bibAnnoteFile{NoStop}{PhysRevLett.85.1146}%
\bibitem{0953-4075-37-5-L01}%
  \BibitemOpen
  \bibfield{author}{%
  \bibinfo {author} {\bibfnamefont{B.}~\bibnamefont{Damski}},\ }%
  \bibfield{journal}{%
  \bibinfo {journal} {Journal of Physics B: Atomic, Molecular and Optical
  Physics}\ }%
  \textbf{\bibinfo {volume} {37}},\ \bibinfo {pages} {L85} (\bibinfo {year}
  {2004})%
  \bibAnnoteFile{NoStop}{0953-4075-37-5-L01}%
\bibitem{0953-4075-42-17-175302}%
  \BibitemOpen
  \bibfield{author}{%
  \bibinfo {author} {\bibfnamefont{B.~B.}\ \bibnamefont{Baizakov}}, \bibinfo
  {author} {\bibfnamefont{F.~K.}\ \bibnamefont{Abdullaev}}, \bibinfo {author}
  {\bibfnamefont{B.~A.}\ \bibnamefont{Malomed}},\ and\ \bibinfo {author}
  {\bibfnamefont{M.}~\bibnamefont{Salerno}},\ }%
  \bibfield{journal}{%
  \bibinfo {journal} {Journal of Physics B: Atomic, Molecular and Optical
  Physics}\ }%
  \textbf{\bibinfo {volume} {42}},\ \bibinfo {pages} {175302} (\bibinfo {year}
  {2009})%
  \bibAnnoteFile{NoStop}{0953-4075-42-17-175302}%
\bibitem{leggett2006}%
  \BibitemOpen
  \bibfield{author}{%
  \bibinfo {author} {\bibfnamefont{A.}~\bibnamefont{Leggett}},\ }%
  \emph{\bibinfo {title} {Quantum Liquids}}\ (\bibinfo {publisher} {Oxford
  Univ. Press, Oxford},\ \bibinfo {year} {2006})%
  \bibAnnoteFile{NoStop}{leggett2006}%
\bibitem{PhysRevA.72.023621}%
  \BibitemOpen
  \bibfield{author}{%
  \bibinfo {author} {\bibfnamefont{L.}~\bibnamefont{Salasnich}}, \bibinfo
  {author} {\bibfnamefont{N.}~\bibnamefont{Manini}},\ and\ \bibinfo {author}
  {\bibfnamefont{A.}~\bibnamefont{Parola}},\ }%
  \bibfield{journal}{%
  \Doi{10.1103/PhysRevA.72.023621}{\bibinfo {journal} {Phys. Rev. A}}\ }%
  \textbf{\bibinfo {volume} {72}},\ \bibinfo {pages} {023621} (\bibinfo {year}
  {2005})%
  \bibAnnoteFile{NoStop}{PhysRevA.72.023621}%
\bibitem{PhysRevA.76.015601}%
  \BibitemOpen
  \bibfield{author}{%
  \bibinfo {author} {\bibfnamefont{L.}~\bibnamefont{Salasnich}},\ }%
  \bibfield{journal}{%
  \Doi{10.1103/PhysRevA.76.015601}{\bibinfo {journal} {Phys. Rev. A}}\ }%
  \textbf{\bibinfo {volume} {76}},\ \bibinfo {pages} {015601} (\bibinfo {year}
  {2007})%
  \bibAnnoteFile{NoStop}{PhysRevA.76.015601}%
\bibitem{CompPhysComm.180.1888}%
  \BibitemOpen
  \bibfield{author}{%
  \bibinfo {author} {\bibfnamefont{P.}~\bibnamefont{Muruganandam}}\ and\
  \bibinfo {author} {\bibfnamefont{S.}~\bibnamefont{Adhikari}},\ }%
  \bibfield{journal}{%
  \bibinfo {journal} {Computer Physics Communications}\ }%
  \textbf{\bibinfo {volume} {180}},\ \bibinfo {pages} {1888} (\bibinfo {year}
  {2009})%
  \bibAnnoteFile{NoStop}{CompPhysComm.180.1888}%
\bibitem{rosenbusch2002dynamics}%
  \BibitemOpen
  \bibfield{author}{%
  \bibinfo {author} {\bibfnamefont{P.}~\bibnamefont{Rosenbusch}}, \bibinfo
  {author} {\bibfnamefont{V.}~\bibnamefont{Bretin}},\ and\ \bibinfo {author}
  {\bibfnamefont{J.}~\bibnamefont{Dalibard}},\ }%
  \bibfield{journal}{%
  \bibinfo {journal} {Physical review letters}\ }%
  \textbf{\bibinfo {volume} {89}},\ \bibinfo {pages} {200403} (\bibinfo {year}
  {2002})%
  \bibAnnoteFile{NoStop}{rosenbusch2002dynamics}%
\bibitem{PhysRevA.63.041603}%
  \BibitemOpen
  \bibfield{author}{%
  \bibinfo {author} {\bibfnamefont{J.~J.}\ \bibnamefont{Garcia-Ripoll}}\ and\
  \bibinfo {author} {\bibfnamefont{V.~M.}\ \bibnamefont{Perez-Garcia}},\ }%
  \bibfield{journal}{%
  \Doi{10.1103/PhysRevA.63.041603}{\bibinfo {journal} {Phys. Rev. A}}\ }%
  \textbf{\bibinfo {volume} {63}},\ \bibinfo {pages} {041603} (\bibinfo {year}
  {2001})%
  \bibAnnoteFile{NoStop}{PhysRevA.63.041603}%
\bibitem{PhysRevA.64.053611}%
  \BibitemOpen
  \bibfield{author}{%
  \bibinfo {author} {\bibfnamefont{J.~J.}\ \bibnamefont{Garcia-Ripoll}}\ and\
  \bibinfo {author} {\bibfnamefont{V.~M.}\ \bibnamefont{Perez-Garcia}},\ }%
  \bibfield{journal}{%
  \Doi{10.1103/PhysRevA.64.053611}{\bibinfo {journal} {Phys. Rev. A}}\ }%
  \textbf{\bibinfo {volume} {64}},\ \bibinfo {pages} {053611} (\bibinfo {year}
  {2001})%
  \bibAnnoteFile{NoStop}{PhysRevA.64.053611}%
\bibitem{PhysRevA.64.043611}%
  \BibitemOpen
  \bibfield{author}{%
  \bibinfo {author} {\bibfnamefont{A.}~\bibnamefont{Aftalion}}\ and\ \bibinfo
  {author} {\bibfnamefont{T.}~\bibnamefont{Riviere}},\ }%
  \bibfield{journal}{%
  \Doi{10.1103/PhysRevA.64.043611}{\bibinfo {journal} {Phys. Rev. A}}\ }%
  \textbf{\bibinfo {volume} {64}},\ \bibinfo {pages} {043611} (\bibinfo {year}
  {2001})%
  \bibAnnoteFile{NoStop}{PhysRevA.64.043611}%
\bibitem{PhysRevA.66.023611}%
  \BibitemOpen
  \bibfield{author}{%
  \bibinfo {author} {\bibfnamefont{A.}~\bibnamefont{Aftalion}}\ and\ \bibinfo
  {author} {\bibfnamefont{R.~L.}\ \bibnamefont{Jerrard}},\ }%
  \bibfield{journal}{%
  \Doi{10.1103/PhysRevA.66.023611}{\bibinfo {journal} {Phys. Rev. A}}\ }%
  \textbf{\bibinfo {volume} {66}},\ \bibinfo {pages} {023611} (\bibinfo {year}
  {2002})%
  \bibAnnoteFile{NoStop}{PhysRevA.66.023611}%
\bibitem{PhysRevA.62.063617}%
  \BibitemOpen
  \bibfield{author}{%
  \bibinfo {author} {\bibfnamefont{A.~A.}\ \bibnamefont{Svidzinsky}}\ and\
  \bibinfo {author} {\bibfnamefont{A.~L.}\ \bibnamefont{Fetter}},\ }%
  \bibfield{journal}{%
  \Doi{10.1103/PhysRevA.62.063617}{\bibinfo {journal} {Phys. Rev. A}}\ }%
  \textbf{\bibinfo {volume} {62}},\ \bibinfo {pages} {063617} (\bibinfo {year}
  {2000})%
  \bibAnnoteFile{NoStop}{PhysRevA.62.063617}%
\bibitem{PhysRevLett.86.564}%
  \BibitemOpen
  \bibfield{author}{%
  \bibinfo {author} {\bibfnamefont{D.~L.}\ \bibnamefont{Feder}}, \bibinfo
  {author} {\bibfnamefont{A.~A.}\ \bibnamefont{Svidzinsky}}, \bibinfo {author}
  {\bibfnamefont{A.~L.}\ \bibnamefont{Fetter}},\ and\ \bibinfo {author}
  {\bibfnamefont{C.~W.}\ \bibnamefont{Clark}},\ }%
  \bibfield{journal}{%
  \Doi{10.1103/PhysRevLett.86.564}{\bibinfo {journal} {Phys. Rev. Lett.}}\ }%
  \textbf{\bibinfo {volume} {86}},\ \bibinfo {pages} {564} (\bibinfo {year}
  {2001})%
  \bibAnnoteFile{NoStop}{PhysRevLett.86.564}%
\end{thebibliography}%
\end{document}